# Dynamic interfacial polaron enhanced superconductivity of FeSe/SrTiO$_3$

Shuyuan Zhang,[1,2] Tong Wei,[3] Jiaqi Guan,[1,2] Qing Zhu,[1,2] Wei Qin,[3] Weihua Wang,[1] Jiandi Zhang,[4] E. W. Plummer,[4] Xuetao Zhu,[1,2*] Zhenyu Zhang,[3*] Jiandong Guo[1,2,5*]

[1]*Beijing National Laboratory for Condensed Matter Physics and Institute of Physics, Chinese Academy of Sciences, Beijing 100190, China*

[2]*School of Physical Sciences, University of Chinese Academy of Sciences, Beijing 100049, China*

[3] *International Center for Quantum Design of Functional Materials (ICQD), Hefei National Laboratory for Physical Sciences at Microscale, and Synergetic Innovation Center of Quantum Information and Quantum Physics, University of Science and Technology of China, Hefei, Anhui 230026, China*

[4]*Department of Physics and Astronomy, Louisiana State University, Baton Rouge, LA 70808, USA*

[5]*Collaborative Innovation Center of Quantum Matter, Beijing 100871, China*

**Abstract:**

**The observation of substantially enhanced superconductivity of single-layer FeSe films on SrTiO$_3$ has stimulated intensive research interest. At present, conclusive experimental data on the corresponding electron-boson interaction is still missing. Here we use inelastic electron scattering spectroscopy and angle resolved photoemission spectroscopy to show that the electrons in these systems are dressed by the strongly polarized lattice distortions of the SrTiO$_3$, and the indispensable non-adiabatic nature of such a coupling leads to the formation of dynamic interfacial polarons. Furthermore, the collective motion of the polarons results in a polaronic plasmon mode, which is unambiguously correlated with the surface phonons of SrTiO$_3$ in the presence of the FeSe films. A microscopic model is developed showing that the interfacial polaron-polaron interaction leads to the superconductivity enhancement.**

As one of the most important concepts in many body systems, electron-boson interaction (EBI) determines many fundamental properties of materials. The most prominent example is the formation of Cooper pairs in superconductivity, mediated by a proper type of bosons via EBI. Identification of the precise nature of the mediating bosons is necessarily the key step in understanding any superconducting system, be it conventional or unconventional. Unlike the cases of conventional superconductors where the bosonic modes are phonons associated with lattice vibrations, the mediating modes in unconventional copper- and iron-based superconductors [1,2] are still highly controversial. Even more challenges are posed by the recently discovered interface enhanced superconducting systems of single-layer FeSe films on $SrTiO_3$ (FeSe/STO), where the superconducting transition temperature ($T_C$) can be dramatically enhanced to ~ 30 - 65 K [3-7] (varying due to different measurement methods and definitions) from ~ 8 K in bulk FeSe [8]. Electron doping has been shown to be indispensable in achieving the dramatic interfacial $T_C$ enhancement [7,9]. However, electron doping alone can only enhance $T_C$ up to ~ 40 K regardless of the doping approach [10-16], leaving the extra ~ 25 K enhancement to be inherently tied to the oxide substrates. Furthermore, as for many other unconventional superconductors, what bosonic modes are dominantly mediating and/or enhancing the superconductivity in FeSe/STO has been a perplexing open question.

In a recent study, a specific type of EBI was invoked to be responsible for the $T_C$ enhancement in FeSe/STO, where the electrons in the FeSe are coupled with the phonons in the STO, the latter manifested by the replica bands [17]. However, theoretical model based on the renormalized forward scattering mechanism is controversial [18]. Our recent studies have shown that the electric field associated with the high-energy surface Fuch-Kliewer (F-K) phonons of STO can penetrate into the FeSe films [19,20], making those modes strong candidates for the interfacial EBI. The energy of the involving F-K phonons is much higher than the Fermi energy of single-layer FeSe/STO, so the EBI related to this system is distinct from conventional electron-phonon coupling picture in metals. In this letter, we use complementary studies of both electronic structure and bosonic excitations to reveal the existence of dynamic interfacial polarons through the observation of polaronic plasmons. The polaron-polaron interaction is theoretically shown to be responsible for the superconductivity enhancement.

In this study, single-layer FeSe films were grown on STO (001) substrates by molecular beam epitaxy. The surface bosonic excitations of both single-layer FeSe/STO and bare STO (001) were measured in an angle-resolved high resolution electron energy loss spectroscopy (HREELS) system with two-dimensional (2D) energy and momentum mapping [21]. The electronic structures of single-layer FeSe/STO were determined by angle resolved photoemission spectroscopy (ARPES). Details of the experimental methods are described in the Supplemental Material [22].

Figure 1(a) presents a 2D-HREELS energy-momentum mapping of single-layer FeSe/STO at 35 K. Besides the Fe- and Se-derived phonon branches located within the range of 0 - 40 meV and the F-K phonon modes located at 59 and 97 meV observed previously [19,20], we detect a new mode at the high energy end of 150 meV, which exhibits strong temperature dependence as shown in Fig. 1(b). Here we label the F-K and the new modes as β, α, and ρ, respectively. As we will demonstrate later, this new mode ρ corresponds to the collective excitation of phonon-dressed electrons (or polarons) of the systems, and contains rich information about the interfacial EBI.

We first note that, as a vital reference system, the STO substrate is a polar semiconductor, in which the electrons and their companying lattice polarizations (or phonons) form quasi-particles called polarons [23]. Just like conduction electrons in metals may move collectively as plasmons, the polarons in doped STO [24] can also be excited collectively as polaronic plasmons [25]. For the Nb-doped STO without FeSe, we also observe three modes, located at 59, 97, and 180 meV, as shown in Fig. 1(c). The first two modes are evidently the F-K phonons of β and α, which are preserved and detectable upon FeSe capping [see Fig. 1(a)]. The broad peak ρ' around 180 meV can be identified as the polaronic plasmon mode in the doped system, for two reasons. First, it is energetically close to the reported plasmon mode identified previously using infrared spectroscopy [26-28]. Secondly, ρ' softens upon increasing the temperature [Fig. 1(d)], which can only be attributed to a corresponding increase in the effective mass of the polarons that give rise to the plasmon mode. Details are presented in the Supplemental Material [22].

Returning to FeSe-capped STO, the identification of the ρ mode to be also the polaronic plasmon mode is based on the consideration that this mode exhibits qualitatively similar temperature behaviors as the ρ' mode [Figs. 1(c) and 1(d)]. The STO substrates are always electron doped to realize the necessary charge transfer, and the transferred carriers was shown to accumulate within the first two atomic layers of the FeSe films [29], allowing us to detect the surface plasmon mode on FeSe/STO by the surface sensitive HREELS technique. Consequently the ρ mode is necessarily dominated by the contribution of the polarons in the STO, aside from the contribution of the electrons in the FeSe overlayer. The pronounced energy lowering from ρ' to ρ is consistent with the significant charge transfer upon FeSe capping. The existence of the polaronic plasmon mode is the signature of the effective coupling between the electrons in the FeSe and the phonon modes that dress the electrons in the STO. The essential consequence would be the enhancement of the band effective mass of the FeSe on STO, which is consistent with published data [30]. To probe the interfacial polaron with the plasmon by HREELS enables the identification of the interfacial EBI. Based on the present findings, it is also plausible to make a mechanistic connection between the reported replica bands from the ARPES measurements [17] and the interfacial polarons, an intriguing angle to be exploited by the community in future studies.

The temperature dependence of β, α, and ρ/ρ' modes reveals a striking contrast with and without the FeSe capping, as shown in Figs. 1(e) and 1(f). Without the capping, ρ' softens significantly upon increasing the temperature, but the F-K modes exhibit little change in either the energy or the linewidth, as shown quantitatively in Fig. 1(f) for the α mode and visually in Fig. 1(d) for the β mode. These observations convincingly show that the polaronic plasmon mode is not effectively coupled with the F-K phonons on clean STO. In contrast, with the FeSe capping, both the ρ and α modes show pronounced softening upon increasing the temperature in a highly correlated manner, potentially also synchronized, an intriguing observation reported in a recent time-resolved study of bulk FeSe [31]. It is also observed that the broadening of the α phonon is accordingly significantly enhanced upon FeSe capping, as shown later. Taken together, we reach the conclusion that the EBI of the system has been substantially enhanced upon the FeSe capping by the activated coupling of the plasmon and α phonon.

The underlying reason for such an enhanced EBI could be attributed to the reduced symmetry associated with the capping and charge transfer from the STO into the FeSe. The latter introduces significant variations in the band structure, accordingly in the EBI, and is indispensable for the interfacial enhanced superconductivity [7]. Furthermore, the electric field associated with the F-K phonons has been shown to be able to penetrate into the FeSe films [19]. Consequently, the transferred electrons in the FeSe and F-K phonons associated with the STO are expected to have enhanced coupling at the interface with spatial confinement and lowered symmetry. Considering that the FeSe electrons participating in the superconductivity are also part of the electron gas complex that gives rise to the polaronic plasmon, the two collective phenomena of superconductivity and polaronic plasmon oscillations are inherently correlated.

Now that the α mode has been recognized to be a major component of the interfacial EBI, it becomes highly desirable to reveal the microscopic nature of the electron scattering processes activated by this mode. Such information is extracted from the electronic structures of the single-layer FeSe/STO system as measured by ARPES shown in Fig. 2(a) and the phonon dispersions from HREELS shown earlier in Fig. 1(a). From energy conservation constraints, we expect that both intra-band electron scatterings within the μ band and inter-band scatterings from μ to φ and from ν to μ can take place, all mediated by the α phonon, but with smaller or larger momentum transfer, respectively. The corresponding scattering magnitudes in the phonon momentum space, shown in Fig. 2(b), demonstrate that both intra- and inter-band transitions are strong. Here the scattering magnitudes are illustrated by the imaginary part of the Lindhard response function $\chi(\mathbf{q}, \hbar\omega_B)$, where $\hbar\omega_B \sim 97$ meV is the energy and $\mathbf{q}$ the momentum of the α phonon. Indeed, as shown in Fig. 2(c), the linewidth of the α phonon is significantly broadened upon the FeSe capping if compared to that without capping, more prominently at larger momenta.

The relative magnitude of the Fermi energy [$E_F \sim 56$ meV, as indicated in Fig. 2(a)] and energy of the α mode ($\hbar\omega_B \sim 97$ meV) renders an important nature of the interfacial EBI: non-adiabaticity. In this regime, the bosons can respond instantaneously to the slowly-moving electrons, as illustrated in Fig. 3a, and contrasted to the adiabatic regime [Fig. 3(b)]. This non-adiabatic nature is also corroborated by the measured electron-momentum relaxation time ($\tau_m \sim 230$ fs) [32], which is much longer than the α phonon pulsation time ($\tau_p \sim 41$ fs). Given the non-adiabatic EBI, the interfacial polarons can migrate in the lattice, and should be described as dynamic interfacial polarons.

The dynamic nature of the interfacial polarons is expected to have important manifestations in mediating the superconductivity if the corresponding EBI is indeed mainly responsible for the enhanced pairing. One of the essential aspects is the short coherence length, corresponding to local pairing [33]. In the single-layer FeSe/STO, the coherence length of a Cooper pair is estimated to be $\xi_{pair} = \hbar v_F / \pi\Delta(0) \sim 1.2$ nm, where $v_F = 1.12 \times 10^5$ m/s is the Fermi velocity and $\Delta(0) = 20$ meV is the superconducting gap extracted from Fig. 2(a). The distinctly short coherence length determined here is also consistent with that of existing vortex radius measurements [3], which is much smaller than the value (~5 nm) of bulk FeSe [34].

To further exploit the local pairing nature of the dynamic interfacial polarons, here we develop a microscopic model to describe the local polaron-polaron coupling, which may lead to superconductivity enhancement. The dynamic nature of the interfacial polarons can induce an additional attractive interaction within a short range (see the Ref. [22] for details). Here the dynamic interfacial polarons only enhance the existing pairing strength in the FeSe without the polar field effect of the STO, irrespective of the detailed pairing mechanism involved. A schematic drawing of the strengthened pairing through the dynamic interfacial polarons is illustrated in Fig. 3(c). Briefly, the lattice distortions associated with the α mode can respond instantaneously to the locally paired electrons in the FeSe. Correspondingly, those Cooper pairs acquire an additional local attractive interaction, as enhanced by the penetrating polarized field of the substrate. Quantitatively, the polaron-polaron attraction $V_{pp}$ is calculated to be $V_{pp}$ ~ 40 meV (see details in Ref. [22]), which translates into a $T_C$ enhancement from ~ 40 K for optimally doped FeSe to ~ 65 K. Such an enhancement mechanism is likely to be broadly applicable to other interfacial superconducting systems on polar substrates, such as STO, $BaTiO_3$, and $TiO_2$.

We have shown that the electrons in the single-layer FeSe/STO are dressed by the strongly polarized and penetrating electric field associated with the phonons of the substrate, leading to the formation of dynamic interfacial polarons. The corresponding polaronic plasmon has been revealed to be strongly correlated with the surface phonons of the STO upon FeSe capping. We have also developed a microscopic model to show that the dynamic polaron-polaron interaction can induce an additional attraction between the electrons in FeSe, resulting in enhanced superconductivity. The mechanism is expected to be broadly applicable, for example, to other interfacial systems such as FeSe/TiO$_2$ [35,36], FeSe/BaTiO$_3$ [37], FeSe/MgO [38], or other layer-structured crystals with strong polaronic characters. Thus the mechanism may serve as an important guide in discoveries of new superconducting systems with higher $T_C$.

## Acknowledgements

The work was supported by the National Key R&D Program of China (No. 2017YFA0303600), the National Natural Science Foundation of China (No. 11634016), and the Strategic Priority Research Program (B) of the Chinese Academy of Sciences (No. XDB07030100). Z.Z. was partially supported by the National Key R&D Program of China (Grants Nos. 2017YFA0303503 and 2014CB921103) and the National Natural Science Foundation of China (Nos. 11634011, 61434002). J.Z. was partially supported by U.S. NSF through Grant No. DMR 1608865. X.Z. was partially supported by the Youth Innovation Promotion Association of Chinese Academy of Sciences (No. 2016008).

*Corresponding authors: jdguo@iphy.ac.cn; xtzhu@iphy.ac.cn; zhangzy@ustc.edu.cn

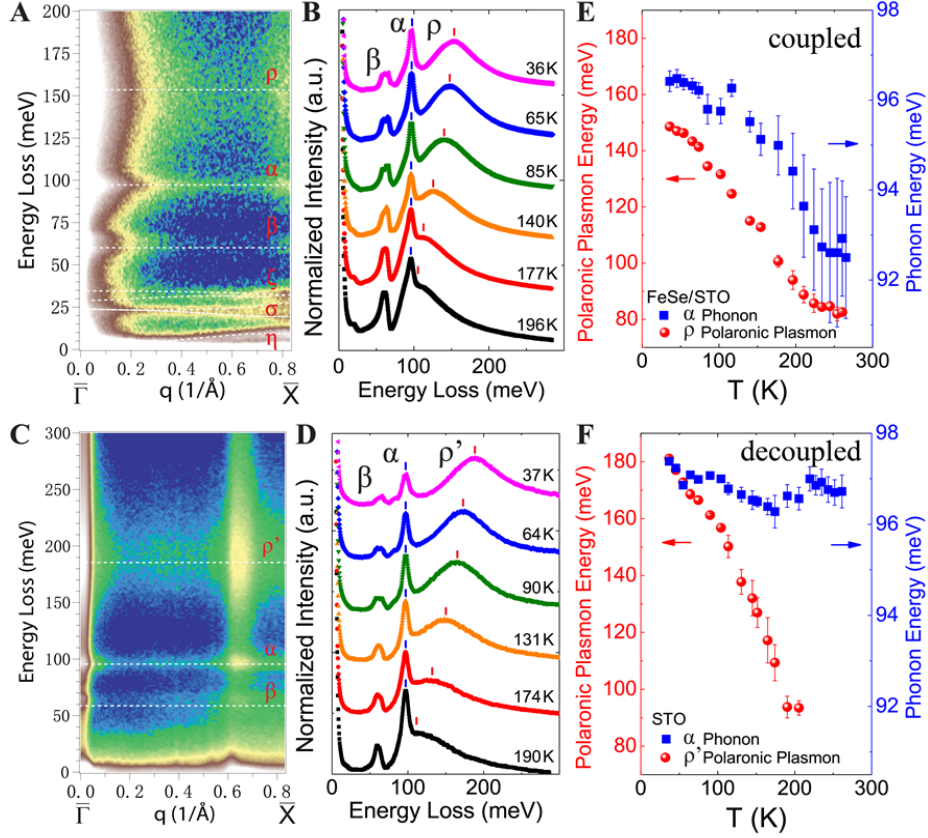

FIG. 1. Phonon and polaronic plasmon modes of the STO with and without FeSe capping. (a) HREELS spectra of the single-layer FeSe/STO measured at 35 K with the incident electron energy of 50 eV. The FeSe phonon branches are labeled by η, σ, and ξ, while the β, α, and ρ modes originated from the STO are described in the main text. (b) Temperature dependence of the energy loss spectra at the $\bar{\Gamma}$ point from the single-layer FeSe/STO. (c) and (d) Corresponding results of the STO without capping, with the incident electron energy of ~ 110 eV. (e) and (f) Comparisons of the energies of the polaronic plasmon and α mode at $\bar{\Gamma}$ point at different temperatures with and without FeSe, respectively. The error bars are from different fitting functions and different background subtraction methods used in the peak fitting.

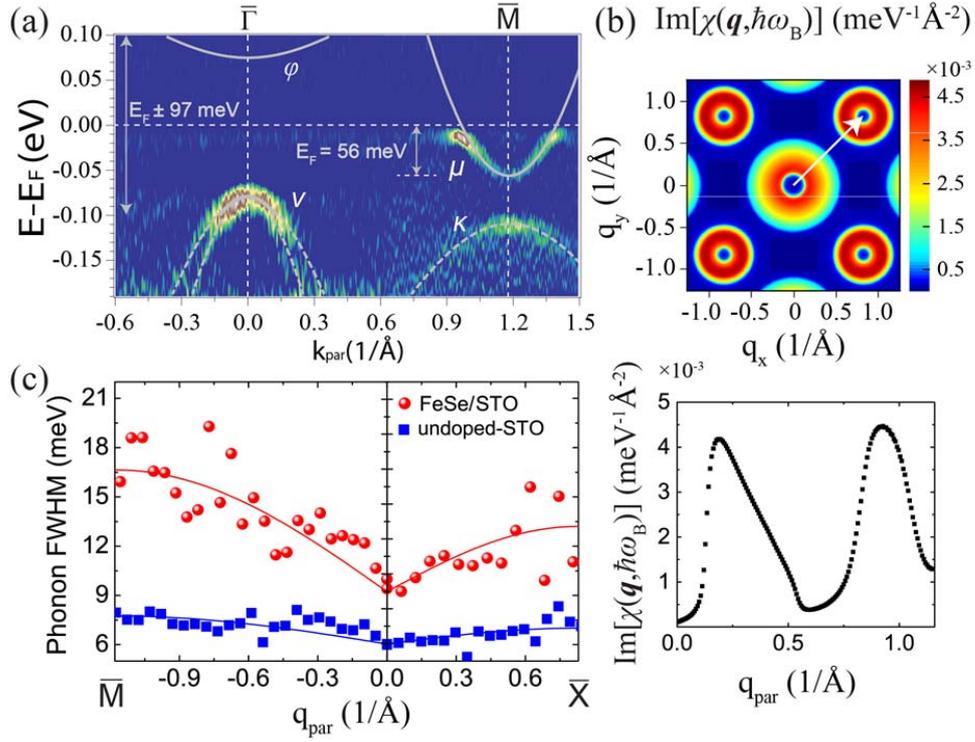

Fig. 2. Electronic band structure and characteristics of the interfacial EBI. (a) Band structure of single-layer FeSe/STO measured at 35 K by ARPES, with the bands above the Fermi energy extracted from Ref [39]. The solid and dashed lines are guides to the eye. (b) Imaginary part of the Lindhard response function, Im[$\chi(q, \hbar\omega_B)$], calculated from the band structure in (a) and dispersion of the α phonon in Fig. 1(a), see the main text for details. The lower panel displays the line profile along the arrow indicated in the upper panel. (c) Full width at half maximum (FWHM) of the α phonon of single-layer FeSe/STO and undoped-STO at 35 K.

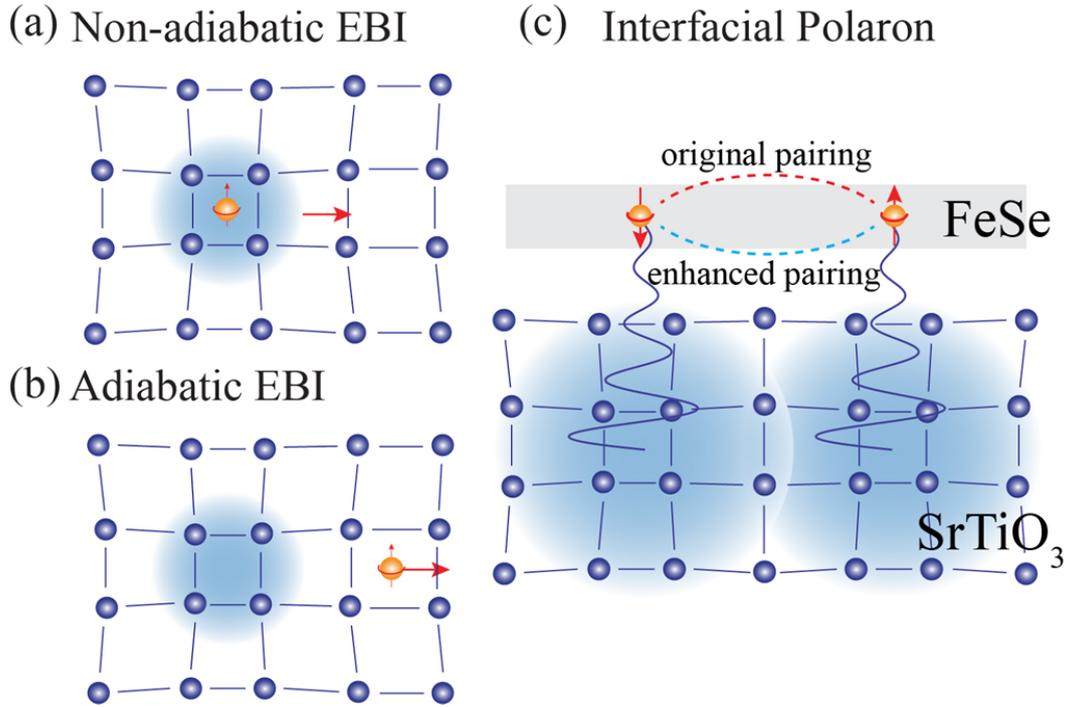

FIG. 3. Illustrations of the non-adiabatic EBI and strengthened pairing. (a) Non-adiabatic and (b) adiabatic EBI, with the size of polaron highlighted by the shaded areas. (c) Enhanced pairing of the dynamic interfacial polarons in single-layer FeSe/STO. The yellow solid spheres indicate the electrons. The curved lines indicate the penetrating polarized field from the STO into the FeSe. The original and enhanced pairing of the electrons are represented by the red and blue dashed lines, respectively.